# NeckCheck: Predicting Neck Strain using Head Tracker Sensors

BHAWANA CHHAGLANI, University of Massachusetts Amherst, USA

ALAN SEEFELDT, Dolby Laboratories, USA

Tech neck, a growing musculoskeletal concern caused by prolonged poor posture during device use, has significant health implications. This study investigates the relationship between head posture and muscular activity in the upper trapezius muscle to predict muscle strain by leveraging data from EMG sensors and head trackers. We train a regression model to predict EMG envelope readings using head movement data. We conduct preliminary experiments involving various postures to explore the correlation between these modalities and assess the feasibility of predicting muscle strain using head worn sensors. We discuss the key research challenges in sensing and predicting muscle fatigue. The results highlight the potential of this approach in real-time ergonomic feedback systems, contributing to the prevention and management of tech neck.

## 1 Introduction

The rise of technology has led to an increase in health issues related to prolonged device usage, such as eye strain, carpal tunnel syndrome, and musculoskeletal disorders. Among these, tech neck has become a major concern, affecting individuals who spend long hours using smartphones, tablets, and computers [1]. Tech neck is characterized by chronic neck pain, stiffness, and musculoskeletal strain resulting from the forward head posture (FHP) commonly observed in device users. This condition places excessive stress on the cervical spine and associated muscles, including the upper trapezius, leading to long-term complications such as cervical spondylosis, disc herniation, and occipital neuralgia. With approximately 75% of the global population spending hours daily with their heads flexed forward [6], there is an urgent need for effective solutions to mitigate this issue. Given the growing prevalence of tech neck and its associated health risks, it is crucial to develop methods for sensing and monitoring this condition in real time to enable early intervention and prevent long-term damage. Several researchers have explored posture sensing techniques to detect poor posture and provide corrective alerts [2, 5, 7]. These methods typically rely on wearable sensors, vision-based systems, or smartphone-based applications to classify good and bad posture. Prior work has focused on real-time posture monitoring and generating alerts to encourage users to maintain ergonomic positions [8]. However, these approaches do not capture the physiological impact of prolonged poor posture, such as muscle strain and fatigue. Additionally, posture correction alerts may be ignored or may not effectively prevent muscle fatigue. If we could predict muscle strain directly instead of relying solely on posture sensing, we could intervene before the user experiences pain, potentially preventing chronic musculoskeletal disorders. This can lead to effective long term assessments which can help individuals in understanding what activities are major contributors of neck strain.

In this work, we study the feasibility of sensing muscle strain using existing head tracker sensors. We observe muscle activity using EMG and use roll, pitch, and yaw data from the head tracker sensor. We study the correlation between muscle activity and head tracker data for different postures associated with tech neck like neck bend, hunching, and FHP. By training regression models to predict EMG envelope readings from head tracker data, we achieve high prediction accuracy, obtaining an R² value of 0.97 and a mean squared error (MSE) of 1.9 using a random forest

---

regression model. These results indicate that neck muscle activity can be effectively predicted using head movement data. Additionally, we discuss key challenges and potential research directions in predicting neck strain. First, extracting muscle strain information from EMG signals is inherently difficult. Second, collecting ground truth data for muscle strain is challenging, as back muscles, unlike biceps, do not easily exhibit strain. Thus, establishing a reliable ground truth for validation remains difficult. Additionally, EMG signals are highly susceptible to noise and can be affected by slight changes in electrode positioning, hair, and other external factors. Addressing these challenges is essential to developing robust and reliable muscle strain monitoring systems. We aim to address these challenges and design an effective system that can predict back muscle fatigue.

## 2 NeckCheck Design

**Sensor Placement**: To capture muscle activity, we placed the EMG sensor on the upper trapezius muscle at the cervical spine. The reference electrode was positioned on the bone on the back to ensure minimal interference from muscle signals and obtain a stable baseline measurement. The IN- and IN+ electrodes were placed close to each other in both placements, as shown in Figure 2. The upper trapezius muscle was chosen for EMG sensing as it is subconsciously activated during postural adjustments, making it an effective region for detecting muscle strain [3]. The head tracker sensor was securely attached to a headphone to track head movements accurately without obstructing natural motion. We collected data simultaneously from both sensors, with the head tracker data sampled at 50 Hz and the EMG data recorded at 500 Hz frequency. Later, we perform timestamp based matching to get both sensor streams synchronized.

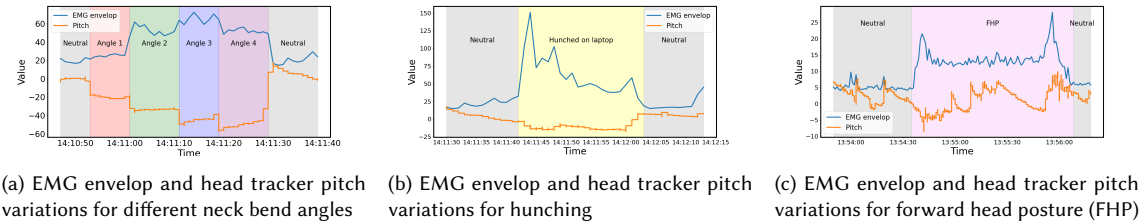

(a) EMG envelop and head tracker pitch variations for different neck bend angles

(b) EMG envelop and head tracker pitch variations for hunching

(c) EMG envelop and head tracker pitch variations for forward head posture (FHP)

Fig. 1. Muscle activity and head movement data for different postures

**Important Postures for Tech Neck:** Tech neck manifests in various postural habits that contribute to muscle strain and fatigue. We identified the following key postures:

*Neck Bend*: Different angles of neck bending were examined to analyze their effect on muscle activation (Angle 1< Angle 2< Angle 3< Angle 4) as shown in Figure 1a. Increasing the neck flexion resulted in progressively higher EMG envelope values, indicating greater muscle strain. Slight neck bend (Angle 1) did not increase EMG significantly, but higher neck bend angles cause increase in EMG. After a certain angle (> Angle 3), EMG remained the same.

*Hunching*: A common posture observed when using laptops or mobile devices, hunching places increased stress on the trapezius muscle. Figure 1b illustrates the variations in EMG signals corresponding to a hunched posture, showing increased muscle activation compared to a neutral position. This confirms that hunching places a high load on neck muscles, leading to potential fatigue and discomfort.

*Forward Head Posture (FHP)*: FHP is one of the most significant contributors to tech neck, where the head is positioned forward relative to the shoulders. As shown in Figure 1c, this posture leads to sustained muscle activity, highlighting the chronic strain induced by prolonged FHP. The EMG envelope remains consistently elevated throughout the FHP phase, indicating sustained muscle engagement. This prolonged strain suggests that FHP contributes to chronic muscle fatigue due to the persistent forward tilt of the head.

These results confirm that poor posture patterns contribute to muscle strain and suggest that head tracker pitch data can serve as a reliable predictor of muscle activity, enabling proactive interventions to mitigate tech neck.

**Predicting EMG using head tracker data** To estimate muscle strain, we preprocess EMG signals to extract the envelope, representing the amplitude of muscle activity. Head tracker data, including positional features (tilt, yaw, pitch) were synchronized with EMG readings. We train multiple regression models using head tracker data as input and EMG envelope as the target variable. We achieve good accuracy for predicting muscle activity using head tracker, demonstrating that head movement data can serve as a reliable predictor of neck muscle activity. These findings indicate the potential for real-time, proactive intervention before discomfort or pain occurs, surpassing traditional posture monitoring techniques.

## 3 NeckCheck Evaluation

**Data Collection**: As shown in Figure 2a, for EMG signal acquisition, we use the BioAmp EXG Pill sensor, which records muscle activity. We obtain head movement data using the Supperware Head Tracker, a device equipped with six 3-axis IMUs for high precision and stability (accurate to a fraction of a degree). We record EMG data using an Arduino-based setup, where signals are read from an analog pin at a sampling rate of 500 Hz. A band-pass filter is applied to remove noise and extract the envelope of the signal for further analysis. The head tracker sensor starts transmitting roll, pitch, and yaw data as soon as its application is opened. We read the serial port to capture the data stream using a Python script, which decodes the incoming values and logs them. The head tracker app is also used for calibrating the sensor. We collect data from two participants, each performing designated postures for 7 minutes per posture. We guide the participants through predefined postures critical for tech neck, ensuring consistency across trials.

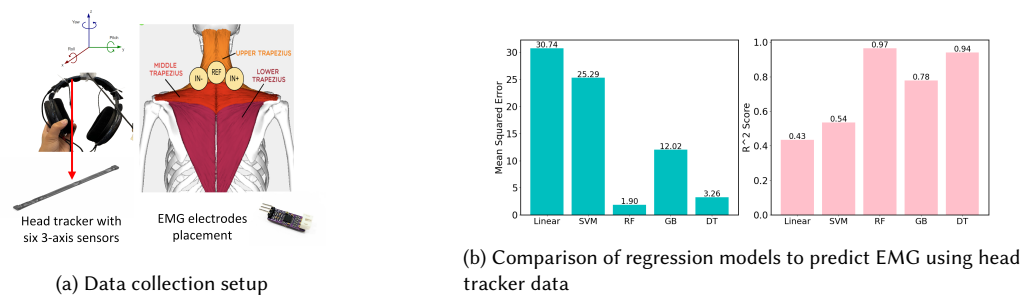

(a) Data collection setup

(b) Comparison of regression models to predict EMG using head tracker data

Fig. 2. Experimental setup and regression results

**Results**: We evaluate multiple regression models to predict EMG envelope readings using head tracker data. Among the models tested, Random Forest Regression (RF) achieves the best performance with an R² score of 0.97 and a Mean Squared Error (MSE) of 1.90. Other models, including Linear Regression, Support Vector Regression (SVR), Gradient Boosting (GB), and Decision Trees (DT), exhibit lower predictive performance, as seen in Figure 2b. RF outperforms all other models by effectively capturing the non-linear relationship between head movement and muscle activity. To further understand the contribution of head tracker features in predicting EMG activity, we analyze feature importance scores from the RF model. Pitch is the most influential feature (0.549), followed by Roll (0.286) and Yaw (0.165). This result suggests that pitch, which corresponds to head tilting forward and backward, has the strongest correlation with muscle strain, making it a crucial factor in detecting tech neck. These results indicate that head movement data can reliably predict muscle strain, enabling real-time interventions before pain or discomfort occurs. By leveraging head tracker sensors, we move beyond posture detection to directly estimating physiological strain, providing a more proactive approach to preventing tech neck-related issues.

## 4  Research Challenges

The results underscore the potential of head tracker data as a non-invasive predictor of muscular activity. This approach could enhance real-time ergonomic feedback systems and contribute to rehabilitation tools that monitor muscle stress. Despite promising results, several challenges must be addressed to improve the accuracy and reliability of predicting muscle strain from head movement data. By addressing these challenges, we aim to improve the robustness and generalizability of our muscle strain prediction model, making it more applicable for real-time ergonomic monitoring and strain prevention.

**EMG Signal Sensitivity and Noise**: EMG signals are highly sensitive to sensor positioning, skin conditions, and external noise. Variations in electrode placement, muscle activity, and hair can significantly affect the readings. To mitigate these issues, we plan to explore advanced signal processing techniques, such as adaptive filtering and machine learning-based noise reduction, to enhance signal stability and reliability.

**Estimating Muscle Fatigue from EMG**: Extracting meaningful muscle fatigue or strain indicators from EMG signals is non-trivial. Current methods rely on observing changes in the center or median frequency of the EMG spectrum as an indicator of muscle fatigue [4]; however, this remains a noisy and imprecise metric. To improve this, we aim to collect data from a larger population and analyze additional features such as signal amplitude variation, power spectral density, and muscle co-activation patterns to develop a more robust metric for muscle strain detection.

**Challenges in Collecting Ground Truth for Muscle Strain**: Unlike biceps, which can be easily fatigued by lifting weights, back muscles are harder to strain under controlled experimental conditions. To address this, we plan to conduct studies involving individuals with pre-existing weaker back muscles or introduce controlled strain conditions, such as carrying a heavy backpack for extended periods, to simulate real-world muscle fatigue scenarios.

## 5  Conclusions

This study demonstrates the feasibility of predicting muscle strain using head tracker data. Our results show that regression models, particularly random forest, can accurately estimate EMG envelope readings, with pitch being the most significant predictor of muscle activity. While challenges remain in reducing EMG noise, refining muscle strain estimation, and improving ground truth collection, future work aims to enhance signal processing techniques and expand study populations. By addressing these limitations, we move closer to developing real-time, non-invasive monitoring systems to prevent tech neck and related musculoskeletal issues.